\documentclass[12pt]{article}
\usepackage{color}
\usepackage{graphicx}
\usepackage[all]{xypic}
\usepackage{psfrag}
\usepackage{tocbibind}
\usepackage{epsf}
\usepackage{epic}
\usepackage{eepic}
\usepackage{epsfig}
\usepackage{wrapfig}
\usepackage{a4}
\usepackage{amssymb,latexsym}
\usepackage[mathscr]{eucal}
\usepackage{url}
\usepackage{fancybox}
\usepackage{ntheorem}
\usepackage{epstopdf}
\usepackage[pdftex,hidelinks]{hyperref}

\newtheorem{theorem}{Theorem}[section]

\newtheorem{lemma}[theorem]{Lemma}

\newcommand{\gnhg}{g}
\newcommand{\mswf}{{\mathsf w}}
\newcommand{\msfw}{{\mswf}}
\newcommand{\Kerr}{\mathrm{Kerr}}

{\catcode `\@=11 \global\let\AddToReset=\@addtoreset}
\AddToReset{figure}{section}

\DeclareFontFamily{OT1}{rsfs}{}
\DeclareFontShape{OT1}{rsfs}{m}{n}{ <-7> rsfs5 <7-10> rsfs7 <10-> rsfs10}{}
\DeclareMathAlphabet{\mycal}{OT1}{rsfs}{m}{n}

{\catcode `\@=11 \global\let\AddToReset=\@addtoreset}
\AddToReset{equation}{section}

\newcounter{mnotecount}[section]

\renewcommand{\themnotecount}{\thesection.\arabic{mnotecount}}

\newcommand{\mnote}[1]%{}
{\protect{\stepcounter{mnotecount}}$^{\mbox{\footnotesize
$%\!\!\!\!\!\!\,
\bullet$\themnotecount}}$ \marginpar{%\color{red}%
\raggedright\tiny\em
$\!\!\!\!\!\!\,\bullet$\themnotecount: #1} }

\newcommand{\jj}[1]%
{{\color{red}\mnote{{\color{red}{\bf jj:} #1} }}}

\definecolor{HP}{rgb}{1,0.09,0.58}

\newcommand{\D}{D}

\newcommand{\eqref}[1]{\eq{#1}}

\newcommand{\hs}{\cH_{\mbox{\scriptsize sing}}}

\newcommand{\beadl}[1]{\begin{deqarr}\label{#1}}
\newcommand{\eeadl}[1]{\arrlabel{#1}\end{deqarr}}%
\def\nz{\ifmmode {I\hskip -3pt N} \else {\hbox {$I\hskip -3pt N$}}\fi}
\def\zz{\ifmmode {Z\hskip -4.8pt Z} \else
       {\hbox {$Z\hskip -4.8pt Z$}}\fi}
\def\qz{\ifmmode {Q\hskip -5.0pt\vrule height6.0pt depth 0pt
       \hskip 6pt} \else {\hbox
       {$Q\hskip -5.0pt\vrule height6.0pt depth 0pt\hskip 6pt$}}\fi}
\def\rz{\ifmmode {I\hskip -3pt R} \else {\hbox {$I\hskip -3pt R$}}\fi}
\def\cz{\ifmmode {C\hskip -4.8pt\vrule height5.8pt\hskip 6.3pt} \else
       {\hbox {$C\hskip -4.8pt\vrule height5.8pt\hskip 6.3pt$}}\fi}
\def\au{{\setbox0=\hbox{\lower1.36775ex\hbox{''}\kern-.05em}\dp0=.36775ex\hs
kip0pt\box0}}
\def\ao{{}\kern-.10em\hbox{``}}

\newcommand\Gregbeq{\begin{eqnarray}}
\newcommand\Gregeeq{\end{eqnarray}}

\def\cH{{\cal H}}

\def\h1{{\hat 1}}
\def\h2{{\hat 2}}

\def\3f{\frac{3}{2}}

\newcommand{\oversetty}[2]{%
\mathop{#2}\limits^{\vbox to -.1ex{%
\kern -1.5ex\hbox{$\scriptstyle #1$}\vss}}}

\newcommand{\jlcax}[1]{}
\newcommand{\eean}{\nonumber\end{eqnarray}}

\newcommand{\kk}[1]{}%{\mnote{{\bf If we consider the KK case:} #1}}

\newcommand{\mcH}{{\mycal H}}

\newcommand{\beq}{\begin{equation}}

\newcommand{\rd}{\,{ d}} % exterior differential

\newcommand{\rgc}[1]{}

\newcommand{\FS}       %{F_1} %
                  {F}
\newcommand{\HS} %{F_2}
       {H_{\mbox{\scriptsize volume}}}

{\ptc{this should be removed in the oberwolfach version}}%

\newcommand{\eel}[1]{\label{#1}\end{equation}}
\newcommand{\eeal}[1]{\label{#1}\end{eqnarray}}
\newcommand{\bed}{\begin{deqarr}}
\newcommand{\eed}{\end{deqarr}}
\newcommand{\bedl}[1]{\begin{deqarr}\label{#1}}
\newcommand{\eedl}[2]{\arrlabel{#1}\label{#2}\end{deqarr}}

\newcommand{\mcU}{{\mycal U}}

\newcommand{\bel}[1]{\begin{equation}\label{#1}}
\newcommand{\bea}{\begin{eqnarray}}
\newcommand{\bean}{\begin{eqnarray}\nonumber}
\newcommand{\beal}[1]{\begin{eqnarray}\label{#1}}
\newcommand{\eea}{\end{eqnarray}}

\newcommand{\nn}{\nonumber}
\newcommand{\Eq}[1]{Equation~\eq{#1}}

\newcommand{\qed}{\hfill $\Box$}
\newcommand{\qedskip}{\hfill $\Box$\medskip}

\newcommand{\be}{\begin{equation}}
\newcommand{\eeq}{\end{equation}}
\newcommand{\ee}{\end{equation}}
\newcommand{\beqa}{\begin{eqnarray}}
\newcommand{\eeqa}{\end{eqnarray}}
\newcommand{\beqan}{\begin{eqnarray*}}
\newcommand{\eeqan}{\end{eqnarray*}}
\newcommand{\ba}{\begin{array}}
\newcommand{\ea}{\end{array}}

\newcommand{\mcD}{{\mycal D}}

\newcommand{\warn}[1]%{}%{}
{\protect{\stepcounter{mnotecount}}$^{\mbox{\footnotesize
$%\!\!\!\!\!\!\,
\bullet$\themnotecount}}$ \marginpar{%\color{red}%
\raggedright\tiny\em
$\!\!\!\!\!\!\,\bullet$\themnotecount: {\bf Warning:} #1} }

\newcommand{\N}{\mathbb N}

\newcommand{\eq}[1]{(\ref{#1})}

\newcommand{\ptc}[1]{\mnote{{\bf ptc:}#1}}

\newcommand{\mcL}{{\mycal L}}

\newcommand{\beqar}{\begin{deqarr}}
\newcommand{\eeqar}{\end{deqarr}}

\newcommand{\beaa}{\begin{eqnarray*}}
\newcommand{\eeaa}{\end{eqnarray*}}

\newcommand{\bethm}{\begin{theorem}}
\newcommand{\et}{\end{theorem}}
\newcommand{\bl}{\begin{Lemma}}

\renewcommand{\D}{{\mcD}}

\begin{document}
\title{Towards a classification of vacuum near-horizons geometries\thanks{Preprint UWThPh-2017-16}}

\author{
    Piotr T.\ Chru\'{s}ciel \\
    Faculty of Physics and Erwin Schr\"odinger Institute\\
    University of Vienna
 \and
 Sebastian J.\ Szybka\\
Jagiellonian Univeristy\\
Copernicus Center for Interdisciplinary Studies
\and
 Paul Tod\\
Mathematical Institute and St.\ John's College, Oxford
}
\date{}

\maketitle

\begin{abstract}
We prove uniqueness of the near-horizon geometries arising from degenerate Kerr  black holes within the collection of nearby vacuum near-horizon geometries.
\end{abstract}

\tableofcontents
\pagebreak

\section{Introduction}

 A lot of effort has been put in general relativity towards classifying all suitably regular stationary solutions of the vacuum equations --- e.g.~\cite{CCH} and references therein. In spite of an impressive body of results, a complete description is still lacking. Indeed, there are unresolved questions concerning analyticity of the solutions, or existence of multi-component solutions, or configurations involving degenerate horizons.

Recall that \emph{degenerate} Killing horizons are, by definition, those Killing horizons on which the surface gravity vanishes. Such horizons arise in an important class of stationary black holes, whose properties are rather distinct from their non-degenerate counterparts. In particular, in vacuum the metric induced by the space-time metric on the sections of degenerate Killing horizons $\mcH$ satisfies the set of equations
\begin{equation}
 \label{31X11.1j}
 R_{AB} -( \mcD _A \omega_B + \mcD _B \omega_A + 2 \omega_A \omega_B) =0
  \;,
\end{equation}
known as the \emph{near-horizon-geometry equations},
where $\omega_A$ is a suitable field of one-forms on the horizon.
The fact that~\cite{Hajicek3Remarks,LP1} (compare  \cite{KunduriLuciettiLR,NurowskiRandall})
\bean &&
 \mbox{all axisymmetric solutions $(g_{AB},\omega_A)$   of \eq{31X11.1j} on a two-dimensional}
\\
&&
 \mbox{{sphere} arise from degenerate Kerr metrics}
\eeal{16III14.1}
plays a key role in the proof that \emph{all connected
degenerate stationary axisymmetric vacuum black holes are Kerr~\cite{ChNguyen}} (see also \cite{FiguerasLucietti,MAKNP}). It is expected that \eq{16III14.1} holds without the axisymmetry assumption.
The goal of this work
is to establish this in a neighborhood of the Kerr near-horizon geometries. Indeed, we prove the following:

\begin{theorem}
  \label{T12VI17.1}
  Let $\cal S$ denote the set of pairs $(g_{AB},\omega_{A})$  on $S^2$  satisfying \eqref{31X11.1j}, let $\cal S_{\Kerr}\subset \cal S$ denote the set of such pairs arising from some Kerr solution. There exists a neighborhood $\mcU$ of $\cal S_{\Kerr}$ in the set of all pairs  $(g_{AB},\omega_{A})$ such that
  $$
  \cal S\cap \mcU = \cal S_{\Kerr}
  \,.
  $$
\end{theorem}

The neighborhood $\mcU$ can be taken, e.g.\ in a $C^2$ topology, or in a $L^2$
topology.

The proof of Theorem~\ref{T12VI17.1} requires controlling the set of zeros of $\omega$ near the Kerr solution. For this we prove quite generally that, on $S^2$, $\omega$ has exactly two zeros, each of index one. We expect this to be useful for a future complete solution of the problem at hand.

Our result suggests the validity of the following: An analytic, vacuum asymptotically flat, suitably regular space-time  with a connected degenerate Killing horizon and with a near-horizon geometry close to Kerr is Kerr. Indeed, one expects that the axi-symmetry of the near-horizon geometry extends to the domain of outer communications by analyticity, so that the usual uniqueness results for connected vacuum black holes~\cite{CCH} apply. However, a proof along these lines would require a careful analysis of isometry-extensions off  degenerate horizons, which does not seem to have been carried out so far.

\section{The Jezierski-Kami\'nski variables}

Let us denote by $(\mathring g,\mathring{\omega})$ the fields describing the near-horizon geometry of  an extreme Kerr metric with mass $m$.
Using a coordinate $x:=\cos\theta$, one has \cite{JezierskiKaminski}:
\begin{equation}\label{gKerr1}
 \mathring{g}=\mathring g_{AB}\rd x^A\rd x^B
 =2m^2\left(a^{-2} \rd x^2 +a^{2}\rd\varphi^2\right)
 \, ,
\end{equation}
\begin{equation}\label{17VI17.5}
 \mathring{\omega}_x=\frac{x}{1+x^2} \, ,\quad
   \mathring{\omega}_\varphi=\frac{a^2}{1+x^2} \, , \quad
% \|\mathring{\omega}\|^2=\frac1{2m^2}\frac{a^2}{1+x^2} \, ,
\end{equation}
where $\displaystyle a^2:= 2\frac{1-x^2}{1+x^2}$.

Invoking the uniformization theorem, metrics on $S^2$ will be described by a conformal factor:
\begin{equation}\label{defu}
    g_{AB}=\exp(2 U)\mathring g_{AB}
     \,.
\end{equation}

There is a `gauge freedom' in $U$ related to the conformal transformations of $(S^2,\mathring g)$, which can be reduced as follows:

Consider any vacuum near-horizon geometry $(S^2,g,\omega)$. By a constant rescaling of $g$ we can, and will, require that the total volume of $g$ equals one. Next,
as shown in section~\ref{s14VI17.2} below,
there exist precisely two points, say $p$ and $q$, on which $\omega$ vanishes. We can, and will, use a conformal transformation of $(S^2,\mathring{g})$ to map $p$ to the north pole and $q$ to the south pole. This leaves the freedom of rotating $S^2$ around the $z$-axis,
as well as performing a conformal transformation generated by the conformal Killing vector
\begin{equation}\label{17VI17.1}
  X:=  a^2 \partial_x
  \,.
\end{equation}

Let $E(g,\omega)$ denote the map which assigns the left-hand side of \eq{31X11.1j} to a metric  $g$ and a one-form field $\omega$.
Let $(\delta g, \delta \omega)\mapsto P(w)$ denote the linearisation of  $E$ at a Kerr solution $(\mathring g,\mathring \omega)$.
We note that the kernel of $P$ is non-trivial, as it contains all deformations of the Kerr near-horizon geometry arising from conformal transformations of $S^2$ \cite{Kaminski}. We will be particularly interested in
\bean
 \mcL_X  \omega
  & = & a^2 \partial_x ( \omega_A dx^A) + 2 a \partial_x a \omega_x dx
\\
 & = &
  \partial_x(a^2 \omega_x) dx + a^2 \partial_x \omega_\varphi d \varphi
  \,.
\eeal{17VI17.2}

Let us denote by ${\mathsf w}= {\mathsf w}_A dx^A$ the linearisation of $\omega$ at the Kerr metric, and by $u$ the linearisation of $U$.
The Jezierski-Kami\'nski variables $(\alpha,\beta)$ are defined as~\cite{JezierskiKaminski}
\begin{equation}
\alpha :=
{\mathsf w}_x -\frac{x}{a^2}{\mathsf w}_\varphi \, ,
\quad
\beta :=
x{\mathsf w}_x +\frac{1}{a^2}{\mathsf w}_\varphi \, ,
%\end{split}
\end{equation}
and it holds that
\begin{equation}
\label{uab}
\partial_A u  = \frac{1}{2}\left[{\mathsf w}_A + \varepsilon_A{^B} \nabla_B \alpha + \nabla_A \beta \right] \, ,
\end{equation}
\begin{equation}\label{newabis}
     {\mathsf w}_x =\frac{\alpha+x\beta}{1+x^2} \, , \quad
    {\mathsf w}_\varphi = a^2\frac{\beta-x\alpha}{1+x^2} \, .
\end{equation}
Suppose that $\omega$ vanishes both at the south and north pole and
let
\bean
 {\mathsf w}_x= \partial_x(a^2 \omega_x)
 %\,,
 %\quad
  \ \mbox{ and } \
 \msfw_\varphi= a^2 \partial_x \omega_\varphi
%  \,.
\eeal{17VI17.3}
be the variations of $\omega$ associated with the conformal Killing vector $X$, as in \eqref{17VI17.2}. At the axes of rotation $x=\pm 1$ the associated functions $\alpha$ and $\beta$ read
\bea
 \alpha= - x \partial_x \omega_\varphi
 \,,
 \quad
 \beta = \partial_x \omega_\varphi
  \,.
\eeal{17VI17.4}
It follows from \eqref{17VI17.5} that $\alpha$ and $\beta$ do not vanish for the Kerr metric on the axes of rotation. This will therefore remain true for all nearby geometries. We conclude that

\begin{lemma}
  \label{l17VI15.1}
  For all near horizon geometries which are $C^1$-close to Kerr, one can find a conformal transformation generated by the conformal Killing vector \eqref{17VI17.1} so that ${\mathsf w}_x$ vanishes at the north pole.
\end{lemma}

This fixes the conformal gauge-freedom up to rotations around the $z$-axis. This remaining freedom turns out to be irrelevant for our purposes.

Let
\[ \triangle_{\mathring g} := \partial_x a^2 \partial_x +
\partial_\varphi a^{-2} \partial_\varphi
%\, .
\]
be the Laplace operator of the metric $\mathring g$.
Set $v:=\left[\begin{array}{c} \alpha \\ \beta \end{array} \right]$,
$\displaystyle B:=\frac1{1+x^2}\left[\begin{array}{cc}
x & -1 \\ 3 & 3x
\end{array}\right]$,
$\displaystyle C:=\frac1{1+x^2}\left[\begin{array}{cc}
1 & x \\ -3x & 3
\end{array}\right]$; then the linearised near-horizon geometry equations
consist of \eqref{uab} together with
\begin{equation}\label{matrixform}
\triangle_{\mathring{g}} v + \frac{4 a^2}{1+x^2}
 \left[\begin{array}{cc}
0 & 0 \\ 0 & 1\end{array}
 \right] v +
\partial_x\left(a^2 B v\right) +
\partial_\varphi\left(C v\right)=0 \, .
\end{equation}

We are ready now to pass to the

\medskip

\noindent{\sc Proof of Theorem~\ref{T12VI17.1}.}
Using the above formulation of the problem we  will show in  section~\ref{s17VI17.1} that the linearised near-horizon geometry operator $P$ has no kernel, once the gauge-freedom inherent in the Jezierski-Kami\'nski formulation has been factored out.

Suppose, for contradiction, that there exists a sequence of pairwise distinct near-horizon geometries $(g_i,\omega_i)$ on $S^2$ converging to $(\mathring g,\mathring \omega)$ as $i\to \infty$. In the conformal gauge just described all metrics have volume one, the zeros of $\omega_i$ are located at the north and south pole, and the associated functions $\alpha_i$ and $\beta_i$ vanish at the north pole.

  Writing $\delta g_i= g_i-\mathring g, \delta \omega_i:= \omega_i - \mathring \omega$, we have
$$
 0 = E(g,\omega)- E(\mathring g,\mathring \omega)= P(\delta g_i, \delta \omega_i) + O(\|\delta g_i, \delta \omega_i\|^2)
 \,.
$$
Dividing by $\|(\delta g_i, \delta \omega_i)\|$ and passing to a subsequence if necessary (here one can invoke elliptic regularity and the Arzela--Ascoli theorem), one finds that $P$ has a non-trivial kernel --- a contradiction. Theorem~\ref{T12VI17.1} follows.
\qedskip

\section{Zeros of $\omega$}
 \label{s14VI17.2}

In this section we show the following:
\begin{theorem}
  Consider a smooth solution of \eqref{31X11.1j} on a two-dimensional sphere $S^2$. Then $\omega=\omega_Adx^A$ has exactly two zeros, each  of index one.
\end{theorem}

Here the index of $\omega$ is understood as that of the associated vector field $\omega^A\partial_A$ obtained by raising indices with the metric.

\medskip

\noindent{\sc Proof.\ } The equation to be solved is
\[
\D_{(A}\omega_{B)}+\omega_A\omega_B=\frac12R_{AB}=\frac14R\gnhg _{AB}.
\]
Prolong by introducing the antisymmetric part of $\D_{A}\omega_{B}$:
\be\label{2}
\D_{[A}\omega_{B]}:=F_{AB}=\frac12\phi\epsilon_{AB},\ee
where $\epsilon_{AB}$ is antisymmetric with $\epsilon_{AB}\epsilon^{AC}=\delta_B^C$  and $\phi$ is real function
global on the sphere. Now the equation can be written as
\be\label{1}
\D_{A}\omega_{B}+\omega_A\omega_B=\frac12K\gnhg _{AB}+\frac12\phi\epsilon_{AB}
 \,,
\ee
where $K=R/2$ is the Gauss curvature.

 Let $x^A$ be normal coordinates centered at a point $p$ where $\omega$ vanishes, oriented consistently with the orientation of $M$. We have
\bean
 |\omega|^2 &= & (-\frac12\phi \epsilon_{AB}
 + \frac K 2 g_{AB})(-\frac12\phi \epsilon_{AC}
  + \frac K 2 g_{AC})x^B x^C + O(|x|^3)
\\
 & = &
\frac 14 \big(\phi(p)^2 + {K(p)^2}\big)|x|^2 + O(|x|^3)
 \;.
\eeal{31I14.2}
If $f(p)^2 + K(p)^2\ne 0$, we see that the zero is isolated. Next, the determinant of
$\mcD _A \omega_B (p)$, which we will denote by $\det$, equals $\frac 14 (\phi(p)^2 + K(p)^2)$. When $\det$ does not vanish the index of $\omega$ at $p$ equals the sign of $\det$, hence one.

Commute derivatives on (\ref{1}) to obtain
\be\label{3}
\D_A\phi+3\omega_A\phi=\epsilon^B_{\;\;\;A}(\nabla_BK+3\omega_BK).\ee

By the uniformisation theorem there is a globally defined, smooth function $u$ (not to be confused with the function $u$ of \eq{uab}) and complex coordinate $\zeta$ with
\be\label{4}
\gnhg _{AB}dx^Adx^B=4e^{2u}\frac{d\zeta d\bar\zeta}{P_0^2},\ee
with $P_0=1+\zeta\bar\zeta$ and $\zeta=\tan\frac{\theta}{2}e^{i\varphi}$.

For later convenience introduce $h^0_{AB}$ for the unit sphere metric, so that
\be\label{44}
h^0_{AB}dx^Adx^B=4\frac{d\zeta d\bar\zeta}{P_0^2}
 \,,
\ee
which is $d\theta^2+\sin^2\theta d\phi^2$, and then $\gnhg _{AB}=e^{2u}h^0_{AB}$.

Introduce the null vector $m^A$ and operator $\delta$ by
\be\label{5}
\delta=m^A\partial_A=\frac{1}{\sqrt{2}}P_0e^{-u}\frac{\partial}{\partial\zeta},\ee
so that also
\[m_Adx^A=\frac{\sqrt{2}e^u}{P_0}d\bar\zeta\mbox{   and  }\gnhg _{AB}=m_A\bar{m}_B+m_B\bar{m}_A,\]
so also
\[
 \epsilon_{AB}=i(m_A\bar{m}_B-m_B\bar{m}_A)
  \,.
\]

We need the Christoffel symbols; we will get them indirectly. Write \mbox{$\delta:=m^A\D_A$} for covariant derivative.
Since $m^A$ is null, there must be complex $\alpha,\beta$  (not to be confused with the $\alpha$, $\beta$ variables of Jezierski and Kami\'nski)  with
\[\delta m^A=\alpha m^A,\;\;\bar{\delta}m^A=\beta m^A
 \,,
\]
whence, by complex conjugation, also
\[
 \bar{\delta}\bar{ m}^A=\bar{\alpha}\bar{ m}^A,\;\;\delta\bar{m}^A=\bar{\beta}\bar{ m}^A
 \,.
\]
Then, since $m_A\bar{m}^A=1$, we deduce $\alpha+\bar{\beta}=0$ (just calculate $\bar{\delta}(m_A\bar{m}^A)$). Finally we can calculate the commutator
\[[\delta,\bar{\delta}]=\bar{\beta}\bar{\delta}-\beta\delta\]
and substitute from (\ref{5}) to deduce that
\be\label{6}\bar{\beta}=-\alpha=\frac{\partial}{\partial\zeta}\left(\frac{1}{\sqrt{2}}P_0e^{-u}\right).\ee
Now we have the connection coefficients explicitly.

\medskip

We proceed by expanding $\omega_A$ in the basis:
\be\label{7}
\omega_A=\chi\bar{m}_A+\bar{\chi}m_A,\ee
for complex function $\chi=\omega_Am^A$,
and then projecting (\ref{1}) along the basis. Contracting with $m^Am^B$ we obtain
\be\label{8}
(\delta-\alpha)\chi+\chi^2=0.\ee
Then with $\bar{m}^Am^B$ to obtain
\be\label{9}
(\bar{\delta}-\beta)\chi+\chi\bar{\chi}=\frac12(K+i\phi)
 \,.
\ee

The one-form $\omega_A$ can be written as a sum of exact and co-exact terms:
\[\omega:=\omega_Adx^A=dv-*dw,\]
where $v,w$ are real-valued, smooth functions on the sphere, unique up to additive constants, and we are using the convention that $(*dw)_A=\epsilon_A^{\;B}w_{,B}$. Then contraction with $m^A$ gives
\be\label{10}\chi=\delta(v+iw)=\delta\psi=\frac{1}{\sqrt{2}}P_0e^{-u}\frac{\partial\psi}{\partial\zeta}\ee
where we have introduced the smooth complex function $\psi=v+iw$, which is unique up to additive complex constant. Using (\ref{6}) rewrite (\ref{8})
as
\[\frac{\partial}{\partial\zeta}\left(\frac{1}{\sqrt{2}}P_0e^{-u}\chi\right)+\chi^2=0,\] and substitute for $\chi$ in terms of $\psi$ to obtain
\[\frac{\partial}{\partial\zeta}\left(\frac{1}{2}P_0^2e^{-2u+\psi}\frac{\partial\psi}{\partial\zeta} \right)=0.\]
This can be integrated in terms of an (at present) arbitrary antiholomorphic function $f(\bar{\zeta})$ as
\[ \frac{1}{2}P_0^2e^{-2u+\psi}\frac{\partial\psi}{\partial\zeta} = f(\bar{\zeta})
 \,,
\]
whence
\be
 \label{11}
 \chi=\frac{1}{\sqrt{2}}P_0e^{-u}\frac{\partial\psi}{\partial\zeta}=\frac{\sqrt{2}}{P_0}e^{u-\psi}f(\bar{\zeta})
  \,.
\ee
We need to constrain $f$. Note that
\[\omega_A\omega^A=2\chi\bar{\chi}=\frac{4}{P_0^2}e^{2u-\psi-\bar{\psi}}f\bar{f}.\]
Since the one-form $\omega_A$ is smooth, it follows that $f$ cannot have singularities in the complex plane of $\zeta$ and must therefore be \emph{entire}. To see what happens
at $\zeta=\infty$ (so to speak) introduce $\eta=-\zeta^{-1}$ and consider
\[\omega=\chi\bar{m}+\bar{\chi}m=2e^{2u}\left(\frac{e^{-\psi}f(\bar{\zeta})d\zeta}{(1+\zeta\bar{\zeta})^2} + \mbox{c.c.}\right)
 \,.
\]
We have
\[
 \frac{f(\bar{\zeta})d\zeta}{(1+\zeta\bar{\zeta})^2}=\frac{\bar{\eta}^2f(-\bar{\eta}^{-1})d\eta}{(1+\eta\bar{\eta})^2}
 \,,
\]
and for boundedness at $\zeta=\infty$, which is $\eta=0$, we need $\bar{\eta}^2f(-\bar{\eta}^{-1})$ bounded there. By an application of Liouville's Theorem
 this forces $f$ to be a quadratic polynomial.
The roots of the quadratic may be distinct or repeated, and they can be moved about by M\"obius transformation, so w.l.o.g.\ there are just two cases to consider:
\begin{enumerate}\item $f=C\bar{\zeta}$;
 \item $f=C\bar{\zeta}^2$;
\end{enumerate}
for complex constant $C$.

\medskip

The next move is to rule out the second case for $f$.

\medskip

Go back to (\ref{3}) and contract with $m^A$ to obtain
\[(\delta+3\chi)(\phi-iK)=0.\]
Substitute for $\chi$ from (\ref{10}) and multiply by $i$ to obtain
\[\delta((K+i\phi)e^{3\psi})=0.\]
Integrate recalling (\ref{5}) to obtain
\[(K+i\phi)e^{3\psi}=g(\bar{\zeta}),\]
for $g$ holomorphic in $\bar\zeta$. This time, the left-hand-side is globally defined on the sphere so that $g$ is a bounded holomorphic function and is therefore a constant, say $C_1$. Thus
\be\label{12}K+i\phi=C_1e^{-3\psi}.\ee
Since $\psi$ is globally defined, (\ref{12}) shows that $K+i\phi$ is everywhere nonzero (if it had a zero then $C_1=0$ so $K=0$ everywhere and we could not be on the sphere), which will give a contradiction to
case 2, as we see next.

\medskip

Go back to (\ref{9}) and substitute for $\chi$ from (\ref{11}). It is clear that, in case 2, all terms on the left vanish at $\zeta=0$, therefore so does $K+i\phi$: contradiction! Thus
\[f=C\bar{\zeta}\]
and
\[\omega_A\omega^A=\frac{4}{P_0^2}e^{2u-\psi-\bar{\psi}}|C|^2\zeta\bar{\zeta}.\]
This vanishes only at $\zeta=0$ in the finite $\zeta$-plane and, as the substitution $\eta=-\zeta^{-1}$ shows, 
at $\eta=0$ (equivalently, $\zeta=\infty$) ---
two isolated simple zeroes which M\"obius transformation places at north and south poles.
\qed

We end this section by noting that it follows from equation \eq{31X11.1j}, together with standard facts about systems of elliptic equations,
that all the fields are real analytic in harmonic coordinates, regardless of the topology of the underlying manifold. This is already clear in any case on $S^2$ from the analysis above.

\section{The kernel of $P$}
 \label{s17VI17.1}

It is shown in \cite{JezierskiKaminski} that elements of the kernel of $P$ are in one-to-one correspondence with solutions of the following system of ODEs on $[-1,1]$ for a sequence of complex functions $(\alpha_k,\beta_k)$, $k\in \N$:
\begin{equation}\label{eqn:final}
 \partial_x( a^2 \partial_x v_k) - \frac{k^2}{a^2} v_k + \frac{4a^2}{1+x^2} \left[ \begin{array}{cc}
 0 & 0 \\
 0 & 1
 \end{array} \right] v_k + \partial_x (a^2Bv_k) + ik(Cv_k) = 0\;,
\end{equation}
where
$$
 v_k=\left[\begin{array}{c} \alpha_k
\\ \beta_k \end{array} \right]
 \,.
$$
The parameter $k$ denotes the $k$th coefficient of $(\alpha,\beta)$ in a Fourier series decomposition with respect to the azimuthal angle $\varphi$ on $S^2$.  It has been proved in \cite{JezierskiKaminski} that for $k\geq 8$
the equation \eqref{eqn:final} does not have solutions  other than $\alpha_k(x)\equiv 0 \equiv \beta_k(x) $, once the relevant boundary conditions, as discussed in the next section, have been imposed. To complete the proof  it remains to prove non-existence of non-zero solutions for the Fourier modes $0< k < 8$, and to analyze the solutions with $k=0$.

\subsection{Boundary conditions}
 \label{s13IV17.1}

Near the north pole $\cos\theta=1$ introduce coordinates $(x^A)=(x^1,x^2)$ defined as
\bel{12VI17.25}
 x^1 = \rho \cos \varphi
 \,,
  \
   x^2 = \rho \sin \varphi
   \,,
   \
   \mbox{with}\
    d\rho^2  =  a^{-2} dx^2
    \,,
    \
    \rho \approx \sqrt{2(1-x)}\ \mbox{for small $\rho$}
    \,.
\ee
The Kerr near-horizon metric is analytic in these coordinates. Analyticity of solutions of systems of elliptic equations implies that  $U$ and $w$ are analytic in these coordinates. A similar construction applies near the south pole.

We have
\bean
 \mswf_A dx^A
  & = &\mswf_x d(\rho \cos \varphi)  + \mswf_y d(\rho \sin \varphi)
  = \frac{\mswf_Ax^A}{\rho} d\rho + \epsilon_{AB}x^A \mswf_B  d\varphi
\\
 &  =
  &
   \frac{\mswf_Ax^A}{a \rho} dx +  \epsilon_{AB}x^A \mswf_B  d\varphi
  \,,
\eeal{12VI17.31}
where $\epsilon_{AB}\in \{0,\pm 1\}$ is totally antisymmetric, and where a sum over $B$ is understood. This gives
\begin{equation}\label{12VI17.31+}
  \alpha =   \frac{\mswf_Ax^A}{a \rho} - \frac{x \epsilon_{AB}x^A \mswf_B}{a^2 }
  \,,
  \quad
  \beta =   \frac{x \mswf_Ax^A}{a \rho} + \frac{  \epsilon_{AB}x^A \mswf_B}{a^2 }
  \,.
\end{equation}
Since $a^2 $ behaves as  $2(1-x)$ i.e.\ as $\rho^2 $ near $\rho=0$, a rough estimate gives \mbox{$\alpha,\beta = O (\rho^{-1})$} there.
However, more can be said if we use a gauge in which $w=0$ at the north and south pole, as can always be done, and which we will assume from now on.
It follows from equation \eq{1} that
\bel{12VI17.24}
 \mswf_B =\frac 12  \left(  \delta \varphi(0) \epsilon_{AB} + {\delta K (0)}   g_{AB}\right)x^A + O(|x|^2)
 \;,
\ee
where $\delta \phi$ and $\delta K$ are the linearised changes of $\phi$ and $M$
associated with the linearised solution. This gives, for small $\rho$,
\begin{equation}\label{12VI17.321}
  \alpha =    \frac{\delta K(0)\rho + O(\rho^2) }{ 2a  } - \frac{  \delta \phi(0)\rho^2
   +O(\rho^3)}{2a^2 }
  \,,
  \quad
  \beta =   \frac{  \delta K(0)\rho + O(\rho^2) }{2a  } + \frac{ \delta \phi(0)\rho^2   +O(\rho^3)}{2 a^2 }
  \,.
\end{equation}
Let $\alpha_k$ and $\beta_k$ denote the $k$th Fourier component of $\alpha$ and $\beta$ in a Fourier series with respect to $\varphi$.
It follows from \eqref{12VI17.321} that
\begin{equation}\label{13VI17.1}
  \alpha_0(0) =  \frac{\delta K(0) -   \delta \phi(0)}{2  }
  \,,
  \quad
  \beta_0(0) =  \frac{\delta K(0)+   \delta \phi(0) }{2  }
  \,.
\end{equation}
However, lemma~\ref{l17VI15.1} shows that we can find a conformal gauge in which $\alpha_0$ and $\beta_0$ vanish at the north pole.

It further follows from what has been said that for $k\ge 1$ we have
\begin{equation}\label{13VI17.2}
  \alpha_k(0) = 0 =
  \beta_k(0)
  \,.
\end{equation}

A similar analysis applies near the south pole $\cos \theta = - 1$.

\subsection{$k=0$}

All solutions of \eqref{eqn:final} with $k=0$ are  found by {\sc Maple} without need of any manipulations of the equations. One obtains
\begin{eqnarray*}
\alpha_0(x)(x^2+1)^2
  &= &C_1\left( x-1 \right)  \left( x+1 \right)\\
&&+ C_2\left( x-1\right)  \left( x+1 \right)  \left[ x+\ln  \left( x-1 \right) -\ln
 \left( x+1 \right)  \right]\\
&&+C_3\, \big\{ [\ln
 \left( x-1 \right)+\ln(x+1)](x^2-1)
 -2 \big\}\\
&&+ C_4  \Big( {\textrm{Li}_2} \left( (x +1)/
2 \right) (x^2-1) +\hat \alpha (x)
  \Big)
\,,
 \end{eqnarray*}
where $\textrm{Li}_2(x)=\sum_{k=1}^\infty\frac{x^k}{k^2}$ is the dilogarithm,  and where $\hat \alpha (x)$ is a lengthy explicit polynomial in $\ln(x-1)$, $\ln (x+1)$ and $x$. Analyticity  of $w_{A}dx^A$ implies that no logarithms or dilogarithms can occur in the solution, hence $C_2=C_3=C_4=0$. Alternatively, a careful analysis of the behaviour of $\alpha$ at $x=\pm1$ together with the requirement of boundedness  leads to the same conclusion. This further results in
\begin{eqnarray*}
 &&\beta_0(x)=-C_1 \frac{x\, \left( {x}^{2}-5 \right)}{(x^2+1)^2}
\,.
\end{eqnarray*}
Translating into $\mswf_Adx^A$, one obtains
$$
 \mswf_x=-\frac{C_1 \left(x^7-6 x^2+1\right)}{\left(x^2+1\right)^3}
 \,,
 \quad
 \mswf_\varphi=-\frac{a^2 C_1 x \left(x^5+x^2-6\right)}{\left(x^2+1\right)^3}
  \,.
$$
Imposing the conformal gauge of lemma~\ref{l17VI15.1} we find $C_1=0$, hence $\alpha_0\equiv0\equiv \beta_0$.

\subsection{$1\le k \le 7$}

When $k\ge 1$, {\sc Maple} finds two explicit linearly independent solutions of \eqref{eqn:final}, the sum of which we denote by $\hat \alpha_k$:
\bean
 \lefteqn{
  \hat \alpha_k(x)(x+i)^2(x-i)^2=
  }
  &&
\\
 \nn
 &&
{ [({x}^2+1)^2-2\,i({x}^{2}-1)/k]}
\\
 &&
 \times
 \left[C_1  \left(\frac{x+1}{x-1}e^{-x}\right)^\frac{k}{2}
  +C_2\left( \frac{x+1}{x-1}e^{-x}\right)^{-\frac{k}{2}}\right]
  \,,
\eeal{s13VI17.11}
where $C_1$ and $C_2$ are arbitrary complex constants.

Now, there is a standard way of obtaining from \eqref{eqn:final} two fourth order decoupled ODEs for $\alpha_k$ and $\beta_k$. Next, there is a standard way of obtaining a lower-order ODE when a solution is known. All this allows one to obtain
$$
 \alpha_k(x)=\hat \alpha_k(x)
  + W_k(x)
 \,,
$$
where $W_k$ solves the following second-order equation
\bel{13VI17.42}
 \left( x-1 \right) ^{2} \left( x+1 \right) ^{2} p_2(x,k) W_k''(x)
 + \left( x+1 \right)
 \left( x-1 \right)  p_1(x,k)  W_k'(x) +p_0(x,k) W_k(x)=0
  \,,
\ee
and where the polynomials $p_i(x,k)$ read
\bean
  p_2(x,k)  & =  &
   8\,  \left( {x}^{2}+1
 \right)  \left[ 1/2\,{k}^{2}{x}^{8}+k \left( i+2\,k \right) {x}^{6}+
 \left( i\,k+3\,{k}^{2}-8 \right) {x}^{4}
 \right.
\\
 \nn
  &&
  \left.
   + \left( -ik+2\,{k}^{2}+16
 \right) {x}^{2} -ik+1/2\,{k}^{2}-8 \right]
 \,,
 \\
 \nn
  p_1(x,k)  & =  &
   16\,
\left[ k \left( i+k \right) {x}^{8}+ \left( 4\,{k
}^{2}-16 \right) {x}^{6}+ \left( -2\,ik+6\,{k}^{2}+64 \right) {x}^{4}
 \right.
\\
 \nn
 &&
  \left.
  +
 \left( 4\,{k}^{2}-80 \right) {x}^{2}+ik+{k}^{2}+32 \right] x
 \,,
\\
 \nn
  p_0(x,k)  & =  &
  -2\Big[ 1/2\,{k}^{4}{
x}^{14}+{k}^{3} \left( i+7/2\,k \right) {x}^{12}+4\,k \left( i{k}^{2}+
{\frac {21\,{k}^{3}}{8}}-6\,i-11\,k \right) {x}^{10}
\\
\nn
 &&
+
  \left( -68\,{k}^
{2}+64+5\,i{k}^{3}+{\frac {35\,{k}^{4}}{2}}-40\,ik \right) {x}^{8}
\\
\nn
&&
+
 \left( 40\,{k}^{2}-128+{\frac {35\,{k}^{4}}{2}}+240\,ik \right) {x}^{
6}
\\
\nn
&&
+ \left( 88\,{k}^{2}+256-5\,i{k}^{3}+21/2\,{k}^{4}+16\,ik \right) {x
}^{4}
\\
 \nn
 &&
+ \left( 4\,{k}^{2}-384-4\,i{k}^{3}+7/2\,{k}^{4}-216\,ik \right)
{x}^{2}
 - \left( {k}^{2}-24 \right)  \left( ik-1/2\,{k}^{2}+8 \right)
 \Big]
  \,.
\eea

\Eq{13VI17.42} is a Fuchsian equation with indicial exponents $\pm k/2$ both at $x=1$ and at $x=-1$.
Near $x=\varepsilon\in\{\pm 1\}$ the solutions $W_k(x)$ have therefore expansions of the form
$$
 W_k(x) = c_{1,\varepsilon,k} {}   (x-\varepsilon )^{-k/2}(P_k{}_{,\varepsilon} {}(x)
  + (x-\varepsilon)^k \varepsilon^k D_k \ln (x-\varepsilon) ) + c_{2,\varepsilon,k} {}    (x-\varepsilon)^{ k/2} +  o (x-1)^{k/2})
 \,,
$$
where $c_{1,\varepsilon,k}$ and $c_{2,\varepsilon,k}$ are  complex constants, the $P_k{}_{,\varepsilon} {}  $'s are polynomials, and the $D_k$'s
can be calculated using {\sc Maple}:
\bean
 \lefteqn{
 \{D_k\}_{k=1}^8
 =
 \bigg\{ -\frac{3}{2} (-1+i),-3,\frac{27 (3+i)}{4},-36 (7+5 i),\frac{225}{8}
   (149+207 i),
   }
   &&
\\
 \nn
  &&
     -243 (225+952 i), \frac{1323}{16} (-33623+127151 i),-1728
   (-302743+270849 i)
    \bigg\}
    \,.
\eea
Since logarithmic terms in $\alpha_k$ are forbidden by the analyticity properties of $w_A dx^A$, we conclude that $c_{1,\varepsilon,k}$ has to vanish for admissible solutions. Thus
\bel{13VI17.44}
 W_k(x) =  c_{2,\varepsilon,k} {}   (x-\varepsilon)^{ k/2} +  o (x-1)^{k/2}
 \,.
\ee
This, together with the requirement of vanishing of $\alpha_k$ at $x=\pm 1$ implies that the constants $C_1$ and $C_2$ in equation \eq{s13VI17.11} vanish, leading to
$$ \alpha_k(x) =   W_k(x)
 \,,
$$
with $W_k$ behaving near $x=\varepsilon$ as in \eqref{13VI17.44}.

\begin{figure}[t!]
\begin{center}
\includegraphics[width=7.5cm,angle=0]{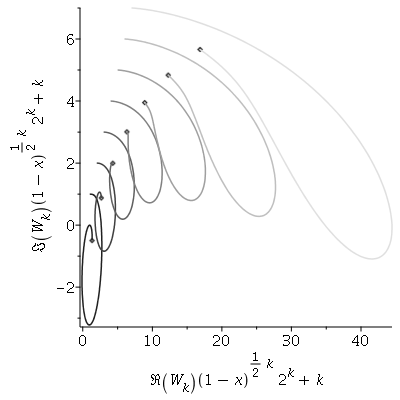}
\caption{\label{F13VI17.1} The curves $[-1,1]\ni x\mapsto 2^{k}(1-x)^{k/2}\big(\Re W_k(x), \Im W_k(x)\big)$ with the boundary condition \eq{13VI17.44+} for $k=1,\dots,7$ (darkness decreasing with $k$), shifted by $(k,k)$ for better readability. The  curves start at $x=-1$, where $W_k$ tends to zero, and approach a non-zero constant in the complex plane as $x$ tends to one, which establishes the $(1-x)^{-k/2}$-behaviour of $W_k$ there.}
\end{center}
\end{figure}

To finish the proof, it remains to show that  the only solution $W_k$ which is regular both at the north and south poles is zero. Assume that this is not the case. Since the problem is linear, there exists a solution which is regular at $x=-1$ (thus $c_1{}_{,-1,k}=0$) with $c_{2,-1,k}=1$.
We have solved   \eqref{13VI17.42}  numerically, using {\sc Maple}\footnote{A limit on the absolute error tolerance for a successful step in the integration must be set carefully.},
 under these conditions,
\bel{13VI17.44+}
 W_k(x) =   (x+1)^{ k/2} +  o( (x+1)^{k/2})
 \,,
\ee
where $o$ is meant for $x$ near $-1$.
Because the equation is singular at $x=-1$,
the numerical solutions have been found by calculating from the equation the first
terms in a power series for $W_k(x)$ (up to order eight, we return to this below), and starting the numerical solution at $x\gtrapprox -1$.
 We then estimated the limit
$$
 \lim_{x\to 1} \, 2^k (1-x)^{k/2}W_k(x)
$$
by stopping the calculation close to $x=1$.
The solutions are plotted in figure~\ref{F13VI17.1}.

It is clear from the figure that all solutions satisfying \eqref{13VI17.44+} blow up at $x=1$ as $(1-x)^{-k/2}$, and thus do not satisfy \eqref{13VI17.44}: Indeed, for these solutions we find  the following numerical estimates
\begin{eqnarray}
 \nn
\{2^k c_{1,1,k}\}_{k=1}^8 &=&\{
0.3293-1.4994i\;, 0.6587-1.1239i\;, 1.3136-1.0047i\;,\\
\nn
&& 2.3401-0.9939i\;,3.9057-1.0507i\;, 6.2775-1.1644i\;,
\\
&& 9.8575-1.3368i\;, 14.1138-1.4603i
\}\;.
\label{17VI17.11}
\end{eqnarray}

Hence
$\alpha_k \equiv 0$ for $k \in \{1,\ldots 7\}$ as well, and thus for all $k\in \N$.

The equality $\beta_k\equiv 0$ directly  follows from this using
\begin{eqnarray}
 \beta_k(x)
  & =  &
2 i \left[\left( -\,{x}^{4}k+2 \left(4 i-k \right) {x}^{2}-8i-k
 \right)  \left( x^2-1 \right)  \left( {x}^{2}+1 \right) W_k'(x)\right.
\\\nonumber
\nn
 &&  \left. +\left(k{x}^{6}+ \left( 8i+9\,k \right) {x}
^{4}+ \left( -16\,i+15\,k \right) {x}^{2}+8i+{7\,k} \right) x W_k(x)
\right]
 \times
\\\nonumber
  &&
 \left[-16\,{x}^{4}+ \left( {x}^{2}+1 \right) ^{4}{k}^{2}+32\,{x}^{2}+2\,i
 \left( x+1 \right)  \left( {x}^{2}+1 \right) ^{2} \left( x-1 \right)
k-16
\right]^{-1}
    \,.
\end{eqnarray}

\begin{figure}[th]
\begin{center}
\includegraphics[height=7cm,angle=0]{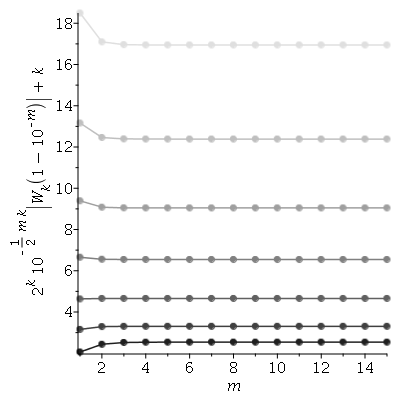}
\end{center}
\caption{\label{F21VI17.1}The value of $2^k 10^{-m\frac{k}{2}}|W_k(1-10^{-m})|$, $k=1,\dots,7$ (darkness decreasing with $k$), near the right end $x=1$, for $m=1,\dots,15$ (the plots are shifted by $k$ for better readability). The initial conditions for the numerical integration were calculated from an expansion of the solution at $x=-1+10^{-m}$.}
\end{figure}

In order to check the convergence of $(1-x)^{k/2}W_k(x)$ as $x$ tends to one, we calculated the values of $2^k 10^{-m\frac k 2}|W_k(1-10^{-m})|$, for $m=1,\dots,15$. The results for $k=1,\dots,7$ are shown in figure~\ref{F21VI17.1}, as calculated using an expansion near $x=-1$ to order eight.

\begin{figure}[th]
\begin{center}
\includegraphics[height=7cm,angle=0]{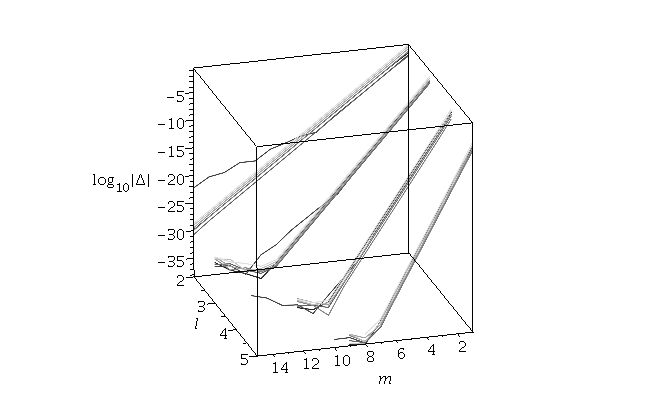}
\end{center}
\caption{\label{F20VI17.1}Convergence of $|2^k c_{1,1,k}|_l$, $k=1,\dots,7$ (darkness decreasing with $k$) as a function of the expansion order $l=2,3,4,5$ at $x=-1+10^{-m}$ compared to $|2^kc_{1,1,k}|_{l=8}$, for $m=1,\dots,15$ ($\Delta=
||2^kc_{1,1,k}|_l-|2^kc_{1,1,k}|_{l=8}|$). The points where $\Delta=0$ are omitted. The points are connected for better readability, with the curves becoming jagged as the limit of numerical precision is reached.}
\end{figure}

Yet another test of the reliability of the results is provided by comparing the values obtained after varying the order of the expansion near $x=-1$. The numerical estimates of $|2^kc_{1,1,k}|$ with $k=1,\dots,7$, calculated using a starting value obtained from an expansion of the solution at  $x=-1+10^{-m}$ for $m=1,\dots,15$ and truncated at order $l=2,3,4,5$, are
compared to $|2^kc_{1,1,k}|$ truncated at order $l=8$ in figure~\ref{F20VI17.1}, where $\Delta=||2^kc_{1,1,k}|_l-|2^kc_{1,1,k}|_{l=8}|$. The points where $\Delta=0$ are omitted.

\bigskip

\noindent{\sc Acknowledgements.} The research of PTC was supported in
part by the Austrian Research Fund (FWF), Project  P23719-N16, and by the Polish National Center of Science (NCN) under grant 2016/21/B/ST1/00940. SJS thanks J.\ Jezierski, L.\ Soko\l owski, Z.\ Golda for a discussion, and acknowledges the support of a grant from the John Templeton Foundation.

\bibliographystyle{amsplain}

\bibliography{}

\end{document}